\documentclass{mem}  

\usepackage{natbib}\usepackage{txfonts}\usepackage{balance}
\usepackage{graphicx}

\begin{document} 
\title{A multi-messenger analysis of neutron star mergers}


\author{
Alessandro Drago\inst{1}, 
Giuseppe Pagliara\inst{1},
Silvia Traversi\inst{1}
          }

\institute{Dip.~di Fisica e Scienze della Terra dell'Universit\`a di Ferrara and INFN Sez.~di Ferrara, Via Saragat 1, I-44100 Ferrara, Italy
}

\authorrunning{Drago}

\titlerunning{Multi-messenger analysis of neutron star mergers}

\abstract{The merger of two neutron stars is a very complex process. In order to disentangle the various steps through which it takes place it is mandatory to examine all the signals we
can detect: gravitational waves and electromagnetic waves, in a 
huge spectrum ranging from X and $\gamma$-rays down to infrared
and radio. Each of these signals provides a message and the totality of this information will allow us not only to understand the process of the merger but also the behavior of
matter at those extreme conditions.

}
\maketitle{}

\section{Introduction}
On August 17, 2017, there has been the first observation of the coalescence of two compact objects with a total mass of the order of 2.74 $M_{\odot}$, which has allowed to identify them as neutron stars (NSs). The source, located at a distance of $40^{+8}_{-8}$ Mpc, has provided different signals: first the gravitational wave (GW170817) signal, detected by the interferometers LIGO and VIRGO; second, with a 1.7 s delay, the Gamma Ray Burst (GRB170817A) and, finally, a bright electromagnetic (EM) counterpart covering all the bands of the spectrum (AT2017gfo). The presence of these different signals regarding the same physical process has determined the beginning of the era of the multi-messenger astronomy \citep{GBM:2017lvd}: indeed there are plenty of informations about cosmology, astrophysics, and nuclear physics which can be inferred by the joined analysis of the whole set of data. In the following we will discuss the physical insides regarding the equation of state (EOS) of neutron stars which can be obtained from the three kinds of signal mentioned before.  

\section{Gravitational waves from the merger of two compact stars}
The calculations of the EOS of the matter composing the inner core of NSs are affected  
by large theoretical uncertaities which translate into large uncertainties regarding the ranges of  masses and radii of NSs.
A possible strategy to pin down the EOS of dense matter consists in measuring both the masses and the radii of the closest NSs (the ongoing NICER experiment is facing this task \citep{Ozel:2015ykl}) but unfortunately radii measurements are affected by large
statistical and sistematic errors \citep{Ozel:2016oaf}. Concerning mass measurements,
the well estabilhed existence of compact stars with masses of $\sim 2M_{\odot}$,
provides strong contraints on the EOS but presently several theoretical possibilities are still viable: the core of NSs could contain just nucleons or it can be partly composed by hyperons \citep{Chatterjee:2015pua} 
and delta resonances \citep{Drago:2014oja}. Also, quark matter can take place in compact stars within hybrid stars (i.e. stars whose core is composed by quark matter) or strange quark stars (i.e. stars entirely composed by strange quark matter). All these possibilities cannot be ruled out by the presently available data. The upcoming measurements of GWs from the merger of two compact stars will surely help 
in reducing the theoretical uncertainties and ultimately to determine the EOS of dense matter.
Let us discuss which will be, in the near future, the phenomenology associated
with the GWs emitted by such systems.
The process of merger of two compact stars can be schematically separated into three main stages:
the inspiral phase, the coalescence phase and the post merger phase; the waveforms of these three stages are qualitatively very different and each of them brings important information 
on the physical parameters of the merger such as the total mass of the system $M$,
the mass asymmetry $q$, the spins of the two components, the orbital parameters, etc.
During the first part of the inspiral phase, when the two stars are at a distance larger than their radii, the GW signal corresponds to the emission of two point-like sources whose orbit is shrinking: both the frequency and the amplitude of the signal increase and their temporal evolution is determined by the so called chirp mass ${\it M}$. This parameter can be measured with high accuracy and it allows to make estimates of $M$ with an error of a few percent.
During the final part of the inspiral phase, the two stars are deformed by the gravitational field of the companion and this leads to a faster evolution towards the coalescence with respect to the case of two point-like sources. This effect, due to the finite size of the two stars, is parametrized by the so called 
tidal parameter $\tilde \Lambda$ which is a function of the tidal deformabilities \citep{Hinderer:2009ca} and the masses of the two stars.
As a general trend: the stiffer the EOS, the larger the value of $\tilde \Lambda$
the stronger are the deviations of the inspiral waveform from the case of point-like sources.
Measuring such deviations would clearly represent a precious constraint on the radii of the two stars.

At the merger, the GW signal reaches its maximum amplitude and frequency. What follows the merger depends on the value of $M$ and on the EoS. 
A first possibility is a prompt collapse to a black hole: in this case the GW signal rapidly switches off, a behavior which is very well characterized in the numerical simulations \citep{Baiotti:2016qnr}. 
Interestingly, the value of the thresholds mass $M_{\mathrm{threshold}}$ above which a prompt collapse takes place can be directly related to the maximum mass of the non-rotating and cold configuration $M_{TOV}$
and on its radius $R_{TOV}$. In \citet{Bauswein:2013jpa,Bauswein:2015vxa,Bauswein:2017aur},
several numerical simulation of the merger, obtained by using different values of $M$ and different 
EOSs have allowed to determine some simple empirical relations between $M_{\mathrm{threshold}}$
and $M_{TOV}$ and $R_{TOV}$ or between $M_{TOV}$ and the radius of the $1.6M_{\odot}$ configuration.
Clearly, a precise determination of $M_{\mathrm{threshold}}$ through GW  measurements 
will constitute a strong contraint on the EOSs. For instance, in the recent 
\citet{Drago:2017bnf}, a strategy has been proposed which will allow 
to test the possible existence of two families of compact stars, hadronic stars and quark stars,
once a few mergers will be detected.

If a prompt collapse does not occur, there are three different types of remnant: a hypermassive star (which is stable as long as differential rotation is present), a supramassive star (which is stable as long as rigid rotation is present) and finally a star which is stable even in absence of rotation. The lifetimes of these different objects are quantitatively very different:
differential rotation can last no longer than about 1 second while rigid rotation can last from $10^4$ s up to millions of years \citep{Falcke:2013xpa}. This implies that while in the first case
no emission of energy is expected at time scales longer than about 1 s, in the other two cases,
the fast rotating star could in principle be the source of powerful electromagnetic emissions as we will discuss in the next section.
During the first hundreds of ms after the merger, the differentially rotating star has a significant quadrupole moment which in turn is responsible for a powerful emission of GWs.
There are several numerical studies of the spectrum of oscillations of the merger remnant 
and the most important modes for what concerns the emission of GWs have been identified, \citep{Bauswein:2011tp,Bauswein:2012ya,Takami:2014zpa,Maione:2017aux}. Detecting such oscillation
modes, would again be important for obtaining information on the EOS: for instance the frequency  of the dominant mode (called $f_{peak}$) scales nicely as the product of $M$ and a quadratic function of $R_{1.6}$, \citep{Bauswein:2012ya}. 
One has to notice however that these modes produce GWs with frequencies above the kHz i.e. in a frequency window for which the sensitivity of the interferometers is low. 

\subsection{GW170817}
Let us finally discuss the first GW event of the era of multi-messenger astronomy: GW170817 \citep{TheLIGOScientific:2017qsa}. The inspiral phase has been clearly detected 
and its signal has allowed to measure a total mass $M \sim 2.74 M_{\odot}$ and to put an upper limit onto 
the value of the tidal parameter $\tilde \Lambda<800$ (with $90\%$ confidence level).
No signal corresponding to the merger and to the following ring-down phase has been detected.
With these two numbers, $M$ and $\tilde \Lambda$ we have already learned something important on the EOS of dense matter: the radius of the $1.4 M_{\odot}$ configuration must be smaller than about $13.4$km (see the analysis of \citet{Annala:2017llu}). Very stiff EoSs, such as MS1 and MS1b which are based on relativistic mean field calculations \citep{Mueller:1996pm} are ruled out. 

The fact that a short GRB without an extended emission has been detected (see discussion in the next section) suggests that the remnant of the merger is a hypermassive star.
In turn this implies that $M_{\mathrm{threshold}}>M$ which translates into a condition on the radius $R_{1.6}>10.7$ km \citep{Bauswein:2017vtn}. Extremely soft EOSs are ruled out by this constraint. Finally, since the remnant most probably is not supramassive, one can 
constrain $M_{TOV}$ to be smaller than about $2.2M_{\odot}$ \citep{Margalit:2017dij,Ruiz:2017due,Rezzolla:2017aly}: EOSs predicting maximum masses larger than this value are therefore also ruled out, e.g. DD2 \citep{Banik:2014qja}.

\section{The associated short GRB}
It is assumed that short GRBs are produced in association with the merger of two neutron stars. Short GRBs can be divided at least in two sub-classes: those
displaying only a prompt emission, whose duration is typically of the order
of a tenth of a second, and those in which an Extended Emission (EE) is observed,
lasting $10^3-10^4$ s and rather similar to the quasi-plateau emission observed
in long GRBs. The problem of finding a mechanism (or maybe the mechanisms) at the
origin of these emissions is therefore two-fold: from one side one has to explain
the duration of the prompt emission, which is two orders of magnitude shorter than
in the case of long GRBs, on the other side one also needs to explain how to generate the EE. 

Concerning the prompt emission, two mechanisms have been proposed. One is based
on the formation of a Black Hole (BH)
\citep{Rezzolla:2011da}: in this case the energy of the GRB is provided by the accretion disk surrounding the BH and the duration of the 
prompt emission is related to the lifetime of the disk.
The other mechanism is based on the formation of a proto-magnetar which
in some 10 s transforms into a Quark Star \citep{Drago:2015qwa}. Here the duration of the prompt
is related to the amount of time during which the ambience surrounding the 
proto-magnetar has the right amount of baryons needed to launch a jet with 
a Lorentz factor of the order of $10^2-10^4$: for a few seconds after the merger
the baryon fraction is too large, while when the process of quark deconfinement
reaches the surface of the proto-magnetar the baryon pollution is strongly suppressed, the Lorentz factor becomes too large and the prompt emission ends.
It is interesting to notice that in both these scenarios the crucial role is played by the  mechanism halting the duration of the prompt emission: in one case the
black hole stops accreting material from the disk, in the other case the formation of the quark star halts the baryon ablation at the origin of the formation of a jet able to generate the prompt emission.

Concerning the EE, a possibility is that it is due to the 
BH accreting mass from a disk \citep{vanPutten:2014kja}. This scenario, although possible in principle, up to now has not been tested on the large set of data
of GRBs with an associated EE. The other possibility is that the
EE is due to the activity of a long living protomagnetar
\citep{Lyons:2009ka,DallOsso:2010uxj,Rowlinson:2013ue},
not collapsing to a BH at least for the duration of the EE, i.e.
$10^3-10^4$s. Let us analyze more in details this second possibility. The protomagnetar scenario needs to be supplemented by the mechanism at the origin of the prompt emission. There are two ways to combine prompt emission and EE: either the prompt
is due to the formation of a BH or is due to the formation of a 
strange quark star. In the first case the so-called "time-reversal" mechanism is needed \citep{Rezzolla:2014nva,Ciolfi:2014yla}, in which the EE is associated
with the activity of the protomagnetar taking place before the
collapse to a BH, but it appears after the prompt emission because it needs to leak through the thick cocoon surrounding the collapsing object. In the second case the prompt emission does take place while the the protomagnetar converts from hadrons to quarks and the EE is due associated with a strange quarks star acting as a magnetar: no "time-reversal" is needed in this second case \citep{Drago:2015qwa}. 

It is interesting to notice that the two scenarios describing the EE as due to a protomagnetar can easily be distinguished by future
observations. In the case of the "time-reversal" scenario the
collapse to a BH, associated with the prompt emission (observed in x and $\gamma$-rays), takes
place at least $10^3-10^4$ s after the moment of the merger (observed in gravitational waves), while in the strange quark star
scenario the prompt emission takes place about 10s after the merger (time needed for the quark deconfinement front to reach
the surface of the star \citep{Drago:2015fpa}). This is a typical example of the way a multi-messenger analysis can discriminate among different possible mechanisms.

\subsection{GRB170817A}
In the case of the event of August 2017 the GW signal clearly indicates that a merger did take place but, on the other hand, the $\gamma$-ray emission was delayed by approximately two seconds respect to the moment of the merger and the observed signal was much weaker than the one of a typical short GRB. It is also relevant to stress that no extended emission was observed, likely indicating that a supramassive star did not form after the merger.

There are two main possible interpretations of the event. The first one assumes that the emission was intrinsically sub-luminous and quasi-isotropic 
\citep{Gottlieb:2017pju,Kasliwal:2017ngb}. The second one assumes instead a standard short GRB emission, that was observed off-axis \citep{Lazzati:2017zsj}. While at the moment, about a hundred days after the event, both possibilities can explain the data, the analysis of the future time-evolution of the emission will ultimately be able to distinguish between these two scenarios, telling therefore if GRB170817A was a standard short GRB seen off-axis or if it belongs to a new class of phenomena \citep{Margutti:2018xqd}.

Even though at the moment the mechanism which launched GRB190817A is still unclear, some strongly energetic emission in $\gamma$ and in x-rays was produced and this indicates that the merger did not collapse instantaneously to a BH. There are explicit simulations indicating that if a jet needs to be formed the object produced in the post-merger needs to survive for at least a few tens of milliseconds \citep{Ruiz:2017inq}. As discussed in the following, also the analysis of the kilonova emission indicates that the result of the merger did not collapse immediately to a BH: a relevant amount of matter was likely emitted from the disk on a time-scale incompatible with an almost instantaneous collapse. This is a very important point to take into account when discussing the possible models for the merger, as we will do in the last section.
\section{The Kilonova signal}
The merger of two neutron stars results in the ejection of part of the mass of the two coalescent objects caused by dynamical, neutrino or viscous driven mechanisms \citep{Hotokezaka:2015eja}. The ejected fluid is reprocessed, undergoing a series of r-processes which allow to synthesize heavy nuclei: the chains of neutron captures, $\beta$-decay, photo-disintegration and fission reactions \citep{Goriely:2011vg,Korobkin:2012uy,Goriely:2013eua,Bauswein:2013yna,Just:2014fka,Siegel:2017jug} are at     
the origin of the EM counterpart of the NS merger event \citep{Metzger:2010sy}. Because of their typical luminosities $\sim 10^{41}-10^{42}$ erg s$^{-1}$, three order of magnitude above a solar mass star Eddington luminosity, these signals are called Kilonovae (KNe).
The features of the KN, in terms of peak timescale, luminosity and effective temperatures, can put some constraints on the parameters characterizing the ejected mass, i.e. the amount of mass $M_{ej}$, the velocity $v$ and the opacity $k$ \citep{Metzger:2010sy}.   

The processes of mass ejection are basically divided in two classes: the dynamical ones, taking place slightly before the merger up to few ms after, and the ejection of part of the matter contained inside the disk formed around the remnant. The last is driven either by neutrinos or by viscous effects and starts about 10 ms after the merger and can last until the eventual collapse of the remnant to a black hole (BH) \citep{Hotokezaka:2015eja}.  

The first component of the dynamical ejecta is the tidal one 
which originates from the deformation of the stars caused by the non-axisymmetric character of the gravitational field. 
The second dynamical mechanism for the mass ejection is the shock that take place at the moment of the merger between the contact surfaces of the two stars and result in the expulsion of part of the crust material \citep{Palenzuela:2015dqa,Sekiguchi:2015dma,Goriely:2015fqa,Bauswein:2013yna}. 

In the next paragraph we will report the features of the recent kilonova detection as the EM counterpart of GW170817 and we will examine the possible connections with the parameters which characterize the different kinds of ejecta and the EOS of the NSs.

\subsection{AT2017gfo}

The spectrum observed on August shows at least two components compatibles with two different KN models: the so called Blue KN, dominant at early time ($\sim$ 1 day after the merger) and the Red KN characterized by longer peak-timescales ($\sim$ a week) \citep{Nicholl:2017ahq,Cowperthwaite:2017dyu} and which has typical wavelengths in the Red and NIR.

The first one is the brighter, with an initial luminosity of the order of $\sim 5\cdot 10^{41}$ erg s$^{-1}$, and it is likely to be associated to a low opacity ejecta, with $k$ spanning between 0.1 cm$^2$s$^{-1}$ and 1 cm$^2$s$^{-1}$ \citep{Roberts:2011xz,Metzger:2014ila}. These values of the opacity characterize a fluid which contains Fe-group or light r-process nuclei, suggesting that the Blue KN is associated with r-processes which synthesize nuclei lying between the first and the second peak i.e. with $A<140$. Moreover, the data analysis indicated an amount of ejected mass ($M^B_{ej}$) of the order of 0.01 $M_{\odot}$ expanding with a velocity of $v^B\sim 0.27$ c \citep{Cowperthwaite:2017dyu,Nicholl:2017ahq}. 

Conversely, the Red KN, which shows a lower luminosity ($\sim 5\cdot 10^{40}$ erg s$^{-1}$), is originated by the r-processes taking place in a fluid with high opacity, from 3 cm$^2$s$^{-1}$ up to 10 cm$^2$s$^{-1}$. This suggests the presence of a relevant ($\sim 10^{-2}$) fraction of Lanthanides, i.e. heavy nuclei with $A>140$, which in turn implies
that elements belongin to third peak of r-processes have been also 
 synthesized. The amount of ejected mass associated with the Red component ($M^R_{ej}$) has been estimated to be $\sim$ 0.04 $M_{\odot}$ and the velocity is relatively low $\sim 0.12 c$ \citep{Cowperthwaite:2017dyu,Chornock:2017sdf}.
 
 The opacity of the ejected fluid can be linked directly to its electron fraction $Y_{e}$: indeed, ejecta characterized by a low $Y_{e}$ are more neutron reach and can, therefore, reach a higher content of Lanthanides elements and correspondingly a higher opacity. This means that the Red KN is likely associated to matter with value of $Y_{e}\sim 0.1-0.2$ while a fluid with $Y_{e}>0.25$ can generate the Blue KN signal.
The electron fraction of the different components of the ejecta depends on the mechanism driving the ejection and on its direction \citep{Wollaeger:2017ahm}. This fact suggests that a detailed analysis of the optical-NIR transient can shed light on the role of the different ejection mechanisms and, as a consequence, on the EOS of the coalescent bodies. 
In particular we can extract different information considering each one of the ejection processes. 

First of all, the tidal ejected mass is characterized by a very low electron fraction $\sim$ 0.1, because the material is mostly ejected in the equatorial plane where the neutrino flux is lower \citep{Hotokezaka:2012ze,Palenzuela:2015dqa,Radice:2016dwd}. This feature reveals that the signal associated to this component is the Red KN \citep{Kasen:2013xka,Barnes:2013wka}: therefore, the large value of $M^R_{ej}$ underlines the importance of this mechanism. This suggests, first of all, that the tidal deformability parameter $\Lambda$ (defined in section 1) cannot be too low in order for the tidal tail to be pronounced and so the amount of mass to be relevant. Since $\Lambda$ is higher for stiffer EOS, extremely soft EOSs seem to appear unfavored. At the same time, a more important tidal effect is associated to a high degree of asymmetry of the binary \citep{Cowperthwaite:2017dyu}. 

For what concerns the shock component, the preferential direction of ejection is the polar one, within an angle of $\sim 30^{\circ}$ \citep{Sekiguchi:2016bjd,Radice:2016dwd}: the more intense neutrino flux causes the raise of the electron fraction up to values $Y_{e}>0.25-0.3$ \citep{Kasen:2013xka,Perego:2017wtu}. The relatively high value of the electron fraction together with the high velocity ($\sim 0.2-0.3 c$) associated with the shock's ejecta allow to identify it as at least one of the components at the origin of the Blue KN. Since the amount of mass ejected by means of this mechanism is proportional to the impact velocity, in order to reach a mass of the ejecta of the order of $M^B_{ej}\sim 10^{-2}\; M_{\odot}$ a soft EOS must be employed. This translates (if the shock provides most of the mass of the Blue KN) in a possible upper limit of the radius of the NS, with a value of $\sim 12$ km, disfavoring the most stiff EOS \citep{Nicholl:2017ahq}. 

Finally concerning the disk's ejecta, the electron fraction of the outgoing fluid is again influenced by the angular distribution: the $Y_{e}$ will be higher for polar ejecta with respect to equatorial one \citep{Perego:2014fma,Fernandez:2013tya,Tanaka:2017lxb,Perego:2014fma,Kasen:2014toa}. Therefore the disk's ejecta can in principle contribute to both the Red and Blue KN. Another important feature influencing the $Y_{e}$ parameter is the lifetime of the remnant: indeed, in the case of a long-lived hypermassive configuration the electron fraction can be raised to an average value of $\sim 0.3 -0.4$, remaining conversely lower for a more prompt collapse \citep{Fujibayashi:2017puw}. This fact determines to which KN component the disk ejecta gives the major contribution. Clearly, the lifetime of the remnant depends again on the stiffness of the EOS. Moreover, the amount of mass potentially ejected is determined by the total mass of the disk (wind and viscosity can drive the ejection of up the 20$\%$ of the disk). To more stiff EOS corresponds a more massive disk because the tidal tales that originate it are more pronounced. 

In conclusion, the measurement of the EM counterpart of NS merger events which can be performed with Theseus \citep{Amati:2017npy,Stratta:2017bwq} can shed light on the importance of the different ejection mechanism and, therefore, to put interesting constraints on the EOS of NS. 

\section{A different hypothesis: a hadronic star - quark star merger}
The event GW170817 and its electromagnetic counterparts have been generated from the coalescence of two compact stars. In the standard scenario, only one family of compact stars does exist, namely the family of stars composed entirely by hadronic degrees of freedom. However, there are some phenomenological indications of the possible existence of a second family of compact stars which are entirely composed by deconfined quarks, namely strange quark stars QSs 
\citep{Drago:2015cea,Drago:2015dea,Wiktorowicz:2017swq}.
In this scenario, the first family is populated by hadronic stars (HSs) which could be very compact and "light" due to the softness of the hadronic EoS (with hyperons and delta resonances included) while the second family is populated by QSs which, on the other hand, can support large masses due to the stiffness of the quark matter EoS.

Within the two-families scenario, a binary system can be composed of two HSs, of two QSs or finally of an HS and a QS.
Let us discuss these three possibilities in connection 
with the phenomenology of GW170817.

The threshold mass $M_{\mathrm{threshold}}$ for a HS - HS, i.e. the limit mass above which a prompt collapse is obtained, has been estimated to be smaller than $\sim 2.7 M_{\odot}$ \citep{Drago:2017bnf}, on the base of the the study performed in \citet{Bauswein:2015vxa}. This value is smaller than the total binary mass $M$ inferred from GW170817 \citep{TheLIGOScientific:2017qsa} and therefore the hypothesis that the binary sytems was a HS-HS system
is disfavored within the two-families scenario.
Also, the possibility that the system was a double QS binary system is excluded because in that case 
it would be difficult to explain the kilonova signal, which is powered by nuclear radioactive decays: even if some material is ejected from the QSs at the moment of the merger, it is not made of ordinary nuclei and therefore it cannot be used inside a r-process chain to produce heavy nuclei.
Conversely, the case of a HS - QS merger, in which the prompt collapse is avoided by the formation of a hypermassive hybrid configuration, becomes the most plausible suggestion in the context of the two-families scenario  \citep{Drago:2017bnf}. 

Under the hypothesis that the event seen in August 2017 is due to the merger of a HS-QS system,
we need now to discuss the possible explanations of the different features seen in the gravitational and electromagnetic signals.
First, the gravitational wave signal has clearly indicated that extremely stiff EoSs are ruled out: the limit put on $\tilde{\Lambda}$ is fulfilled only if the radii of the two stars are smaller than about $13.4$km (see the analysis of \citet{Annala:2017llu}).
Both HSs and QSs satisfy this limit \footnote{A. Drago, G. Pagliara and G. Wiktorowicz, in preparation}\citep{Drago:2013fsa,Drago:2015cea}, see also \citet{Hinderer:2009ca} where the tidal deformabilities of HSs and QSs have been computed. 

Second, the emission of GRB170817A is probably connected with the formation of a relativistic jet which is launched by a BH-accretion torus system. The scenario discussed in \cite{Drago:2015qwa} concerns short GRBs featuring an EE which has not been observed in this case. In our scenario, the compact star which forms immediately after 
the merger is a hypermassive hybrid star in which the burning of hadronic matter is still active. We expect such a system to collapse to a BH once the differential rotation is dissipated. The sGRB would be produced by the same mechanism studied in \citet{Rezzolla:2011da,Ruiz:2016rai}.

Let us finally discuss the properties of the observed kilonova within our scenario. \citet{Perego:2017wtu} suggest  an effective two components model in which the opacity of the secular ejecta is predicted to be very low ($\sim$ 1  cm$^{2}$s$^{-1}$), comparable to that of the wind component. This hypothesis has two major consequences: the lifetime of the remnant must be sufficiently long in order to allow weak reactions to raise the electron fraction to $> 0.3$ and the tidal ejecta must give a very relevant contribution.

Both these requirements can be fulfilled in the context of the HS - QS merger; indeed the hybrid star configuration predicted by this model can survive as a hypermassive configuration for a time of the order of hundreds of ms. Moreover, for an asymmetric binary, characterized by $q=0.75 - 0.8$, the predicted tidal deformability of the lightest star (the hadronic one) can reach value of $\sim 500$. This quite high value of $\Lambda$ together with the supposed high asymmetry of the binary can result in a relevant contribution of the tidal effect to the total ejected mass. This allows to explain the third peak of r-processes and the Red KN without the need of a high opacity secular ejecta (notice also that the value $\Lambda \sim$ 500 is largely above the lower limit derived from the analysis of the EM counterpart performed in \citet{Radice:2017lry}).

In conclusion, despite the need of hydrodynamical simulations in order to make more quantitative predictions, the HS-QS merger can represent a viable way to explain the features of the GW170817, GRB170817A and AT2017gfo. 

\bibliography{biblio}
\bibliographystyle{aa}

\end{document}